\documentclass[aps,prd,preprint,superscriptaddress,tightenlines,nofootinbib]{revtex4}

\usepackage{graphicx}
\usepackage{dcolumn}
\usepackage{bm}

\newcommand{\text}{\mathrm}

\begin{document}

\preprint{CLNS 07/2004}   
\preprint{CLEO 07-8}      

\title{Suppressed Decays of $D^+_s$ Mesons to Two Pseudoscalar Mesons}

\author{G.~S.~Adams}
\author{M.~Anderson}
\author{J.~P.~Cummings}
\author{I.~Danko}
\author{D.~Hu}
\author{B.~Moziak}
\author{J.~Napolitano}
\affiliation{Rensselaer Polytechnic Institute, Troy, New York 12180, USA}
\author{Q.~He}
\author{J.~Insler}
\author{H.~Muramatsu}
\author{C.~S.~Park}
\author{E.~H.~Thorndike}
\author{F.~Yang}
\affiliation{University of Rochester, Rochester, New York 14627, USA}
% ...

\collaboration{CLEO Collaboration} 
\noaffiliation

\date{July 26, 2007}

\begin{abstract} 

Using data collected near the $D^{\ast +}_s D^-_s$ peak production
energy $E_{\text{cm}} = 4170$~MeV
by the CLEO-c detector, we study the decays of $D^+_s$ mesons to
two pseudoscalar mesons. We report on searches for the
singly-Cabibbo-suppressed $D^+_s$ decay modes $K^+ \eta$, $K^+ \eta'$,
$\pi^+ K^0_S$, $K^+ \pi^0$, and the isospin-forbidden decay mode 
$D^+_s \rightarrow \pi^+ \pi^0$. We normalize with respect to the
Cabibbo-favored $D^+_s$ modes $\pi^+ \eta$, $\pi^+ \eta'$, and
$K^+ K^0_S$, and obtain ratios of branching fractions: 
$\mathcal{B}(D^{+}_{s} \rightarrow K^+ \eta)$ / $\mathcal{B}(D^{+}_{s} \rightarrow \pi^+ \eta)$   
= (8.9 $\pm$ 1.5 $\pm$ 0.4)\%,
$\mathcal{B}(D^{+}_{s} \rightarrow K^+ \eta')$ / $\mathcal{B}(D^{+}_{s} \rightarrow \pi^+ \eta')$   
= (4.2 $\pm$ 1.3 $\pm$ 0.3)\%, 
$\mathcal{B}(D^{+}_{s} \rightarrow \pi^+ K^{0}_{S})$ / $\mathcal{B}(D^{+}_{s} \rightarrow K^+ K^{0}_{S})$
= (8.2 $\pm$ 0.9 $\pm$ 0.2)\%, 
$\mathcal{B}(D^{+}_{s} \rightarrow K^+ \pi^0)$ / $\mathcal{B}(D^{+}_{s} \rightarrow K^+ K^{0}_{S})$
= (5.0 $\pm$ 1.2 $\pm$ 0.6)\%, and
$\mathcal{B}(D^{+}_{s} \rightarrow \pi^+ \pi^0)$ / $\mathcal{B}(D^{+}_{s} \rightarrow K^+ K^{0}_{S})$
$< 4.1\% $ at 90\% CL, where
the uncertainties are statistical and systematic, respectively.
\end{abstract}

\maketitle

There are ten possible decays of $D^+_s$ mesons to a pair of mesons
from the lowest-lying pseudoscalar meson nonet. The decay can be to
either $K^+$ or $\pi^+$, combined with any of $\eta$, $\eta'$, $\pi^0$,
$K^0$, or $\bar{K^0}$ ($K^0_S$ or $K^0_L$ for the final
state). 
Measurements of the branching fractions of the complete set of decays
test flavor topology and SU(3) predictions~\cite{RosnerPaper}. 
The Cabibbo-favored, color-favored (external spectator) decays 
$D^+_s \rightarrow \pi^+ \eta$ and $D^+_s \rightarrow \pi^+ \eta'$
have been previously measured~\cite{PDGValue}, 
as has the Cabibbo-favored, color-mixed (internal spectator)
decay $D^+_s \rightarrow K^+ K^0_S$~\cite{PDGValue}. 
Here we present first
observations of the singly-Cabibbo-suppressed, color-favored decays 
$D^+_s \rightarrow K^+ \eta$, $D^+_s \rightarrow K^+ \eta'$, and 
$D^+_s \rightarrow \pi^+ K^0_S$, and strong evidence (4.7 standard
deviations ($\sigma$)) for the singly-Cabibbo-suppressed, color-mixed
decay $D^+_s \rightarrow K^+ \pi^0$. (In this
analysis, we have detected $K^0_S$, but made no attempt to detect
$K^0_L$, nor have previous $D^+_s$ measurements.)
We measure the ratio of the
branching fraction of each singly-Cabibbo-suppressed decay to that of
the corresponding favored decay, expected to be, and found to be, of
order $|V_{cd}/V_{cs}|^{2} \approx 1 /20$. The decay 
$D^+_s \rightarrow \pi^+ \pi^0$ requires a change in isospin of 2
units, and is thus ``isospin-forbidden'', and expected to be
substantially suppressed. 
Our search for this decay reveals no
firm evidence for it, and we present an upper limit.

CLEO-c is a general-purpose solenoidal detector. The charged
particle tracking system covers a solid angle of 93\% of $4 \pi$
and consists of a small-radius, six-layer, low-mass, stereo wire
drift chamber, concentric with, and surrounded by, a 47-layer
cylindrical central drift chamber. The chambers operate in a 1.0 T
magnetic field and achieve a momentum resolution of $\sim$0.6\%
at $p=$1~GeV/$c$.  
We utilize two particle identification (PID) devices to separate
charged kaons from pions: the central drift chamber, which provides
measurements of ionization energy loss ($dE/dx$), and, surrounding
this drift chamber, a cylindrical ring-imaging Cherenkov (RICH)
detector, whose active solid angle is 80\% of $4 \pi$.
Detection of neutral pions and eta mesons relies on an
electromagnetic calorimeter consisting of 7784 cesium iodide
crystals and covering 95\% of $4 \pi$. The calorimeter achieves a
photon energy resolution of 2.2\% at $E_\gamma=$1~GeV and 6\% at
100~MeV. The CLEO-c detector is described in detail
elsewhere~\cite{cleo_detector}.

We use 298 $\mathrm{pb}^{-1}$ of data produced in $e^+ e^-$ collisions
using the Cornell Electron Storage Ring (CESR) 
near the center-of-mass energy $\sqrt{s}=4170$~MeV. 
Here the cross-section for the channel of
interest, $D^{\ast +}_s D^-_s$ or $D^+_s D^{\ast -}_s$, is $\sim$1
nb~\cite{DataSample}. 
We select events in which the $D^{\ast}_s$ decays to 
$D_s + \gamma$ (94\% branching fraction~\cite{PDGValue}).
Other charm production totals $\sim$7 nb~\cite{DataSample}, and
the underlying light-quark ``continuum'' is about 12 nb.
We reconstruct $D^+_s$ mesons in all two-body pseudoscalar decay
channels. Throughout this Letter, charge conjugate modes are implicitly
assumed, unless otherwise noted.

We use the reconstructed invariant mass of the $D_s$ candidate,
$M(D_s)$, and the mass recoiling against the $D_s$ candidate, 
$M_\text{recoil}(D_s)
\equiv \sqrt{ (\sqrt{s} - E_{D_s} )^2 - \vec{p}^2_{D_s} }
$, as our primary kinematic variables to
select a $D_s$ candidate. Here $\vec{p}_{D_s}$ is the momentum of
the $D_s$ candidate, 
$E_{D_s} = \sqrt{m^2_{D_s} + \vec{p}^2_{D_s}}$,
and $m_{D_s}$ is
the known $D_s$ mass~\cite{PDGValue}. We make no requirements on the
decay of the other $D_s$ in the event.

There are two components in the recoil mass distribution, a peak
around the $D^\ast_s$ mass if the candidate is due to the primary 
$D_s$ and a rectangular shaped distribution if the candidate is due to the
secondary $D_s$ from $D^\ast_s$ decays. The edges of
$M_\text{recoil}(D_s)$ from the secondary $D_s$ are
kinematically determined (as a function of $\sqrt{s}$ and known masses),
and at $\sqrt{s} = 4170$~MeV, $\Delta M_\text{recoil}(D_s)
\equiv M_\text{recoil}(D_s) - m_{D^\ast_s}$ is in the range
$[-54, 57]$~MeV. Initial state radiation causes a tail on the high
side, above 57~MeV.
We select $D_s$ candidates within the 
$-55~\text{MeV} \le \Delta M_\text{recoil}(D_s) < +55~\text{MeV}$
range.

We also require a photon consistent with coming from 
$D^{\ast +}_s \rightarrow D^+_s \gamma$ decay, by looking at the
mass recoiling against the $D_s$ candidate plus $\gamma$ system,
$M_\text{recoil}(D_s + \gamma)
\equiv \sqrt{ (\sqrt{s} - E_{D_s} - E_\gamma)^2 - (\vec{p}_{D_s} + \vec{p}_\gamma)^2 }$.
For correct combinations, this recoil mass peaks at $m_{D_s}$, 
regardless of
whether the candidate is due to a primary or a secondary $D_s$.  
We require $| M_\text{recoil}(D_s + \gamma) - m_{D_{s}} | <
20~\text{MeV}$. Though there is a 
25\% efficiency loss from this
requirement, it improves the signal to noise 
ratio, important for the suppressed modes.

Our standard final-state particle selection requirements are described
in detail elsewhere~\cite{HadronPaper}.
Charged tracks produced in the $D^+_s$ decay are required
to satisfy criteria based on the track fit quality, 
have momenta above 50~MeV/$c$, and angles with respect to the beam
line, $\theta$, satisfying $|\cos\theta|<0.93$. 
They must also be consistent with coming from the interaction point in
three dimensions. Pion and
kaon candidates are required to have $dE/dx$ measurements within
three standard deviations ($3\sigma$) of the expected value. For
tracks with momenta greater than 700~MeV/$c$, RICH information, if
available, is combined with $dE/dx$.
The efficiencies (95\% or higher) and misidentification rates (a few
per cent) are 
determined with charged pions and kaons from hadronic $D$
decays.

The $K^0_S$ candidates are selected from pairs of oppositely-charged
and vertex-constrained tracks having invariant mass within
12~MeV, or roughly 4.5$\sigma$, of the known $K^0_S$ mass.
We identify $\pi^{0}$ candidates via $\pi^{0} \rightarrow \gamma \gamma$,
detecting the photons in the CsI calorimeter. 
To avoid having both photons in a region of poorer energy resolution,
we require that at least one of the photons be in the ``good barrel''
region, $|\cos \theta_{\gamma}| < 0.8$. We require that the
calorimeter clusters have a measured energy above 30~MeV, have a
lateral distribution consistent with that from photons, and not be
matched to any charged track.
The invariant mass of the photon pair is
required to be within 3$\sigma$ ($\sigma\sim$ 6 MeV) of the
known $\pi^0$ mass. A $\pi^0$ mass constraint is imposed when $\pi^0$
candidates are used in further reconstruction.
We reconstruct $\eta$ candidates in two decay modes. For the decay
$\eta \rightarrow \gamma \gamma$, candidates are formed using a similar
procedure as for $\pi^0$ except that $\sigma\sim$ 12~MeV. For
$\eta \rightarrow \pi^+ \pi^- \pi^0$, we require that the invariant mass
of the three pions be within 10~MeV of the known $\eta$ mass. We
reconstruct $\eta'$ candidates in the decay mode 
$\eta' \rightarrow \pi^+ \pi^- \eta $. We
require $|m_{\pi^+ \pi^- \eta} - m_{\eta'}|< 10$~MeV.

The $D_s$ invariant mass distributions of the backgrounds to 
$D^{+}_{s} \rightarrow K^+ K^{0}_{S}$ and 
$D^{+}_{s} \rightarrow \pi^+ K^{0}_{S}$ are not smooth, but have
bumps, caused by $D^{\ast +} D^{\ast -}$ events followed by
$D^{\ast \pm} \rightarrow \pi^{\pm} D^0$ decays. 
The low-momentum $\pi^{\pm}$ from $D^{\ast \pm}$ decay, in combination with a
particle from $D^0$ decay, can create a fake $K^0_S$.
To reduce the bump
structure, which complicates fitting the background, we reject those
$D^{+}_{s} \rightarrow K^+ K^{0}_{S}$ and 
$D^{+}_{s} \rightarrow \pi^+ K^{0}_{S}$ candidates that contain a
$\pi^+$ or $\pi^-$ with momentum below 100~MeV/$c$. 
For symmetry, we also reject events with a $K^{\pm}$ with momentum
below 100~MeV/$c$. Further,
we require that the $K^0_S$ has traveled a measurable distance from
the interaction point before decaying, {\it i.e.}, that the distance
along the flight path, from interaction point to $K^0_S$ decay vertex,
be greater than zero with a 3$\sigma$ significance. After the
low-momentum 
track veto and $K^0_S$ flight significance requirement are applied, no
bump structures remain.

For the modes with $\eta$ or $\eta'$, $\eta \rightarrow \gamma \gamma$, 
we reject the $\eta$ candidate if either of the daughter photons is
consistent with coming from $\pi^0 \rightarrow \gamma \gamma$ when
paired with any other $\gamma$ in the event.
This veto reduces the background from fake $\gamma \gamma$ combination
for $\eta$ candidates.

\begin{figure}
  \includegraphics*[width=3.3in]{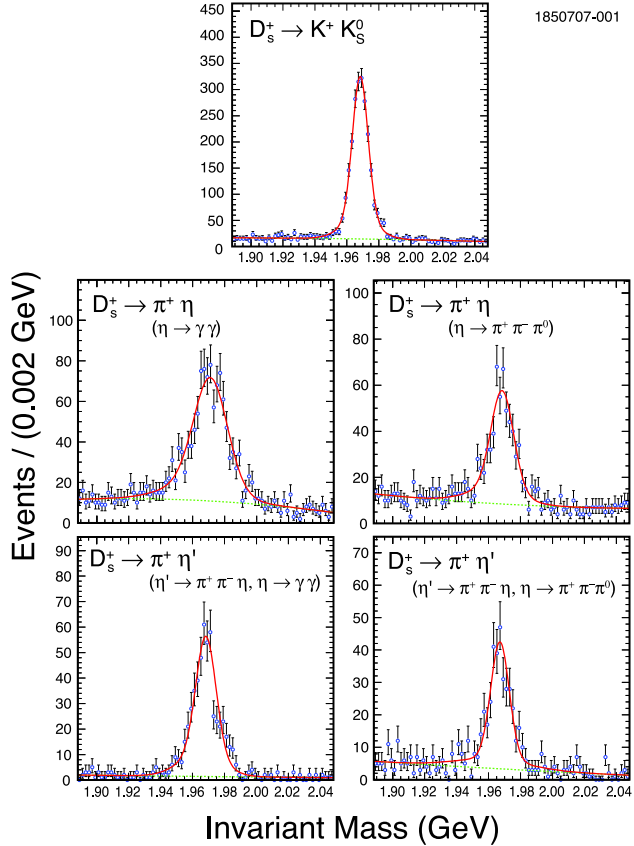}\\
  \caption{$M(D_s)$ distributions for Cabibbo-favored $D_s$ modes from
    data. The points are the data and the superimposed line is the
    fit (the dotted line is the fitted background) 
    as described in the text.} 
  \label{fig:01}
\end{figure}

\begin{figure}
  \includegraphics*[width=3.3in]{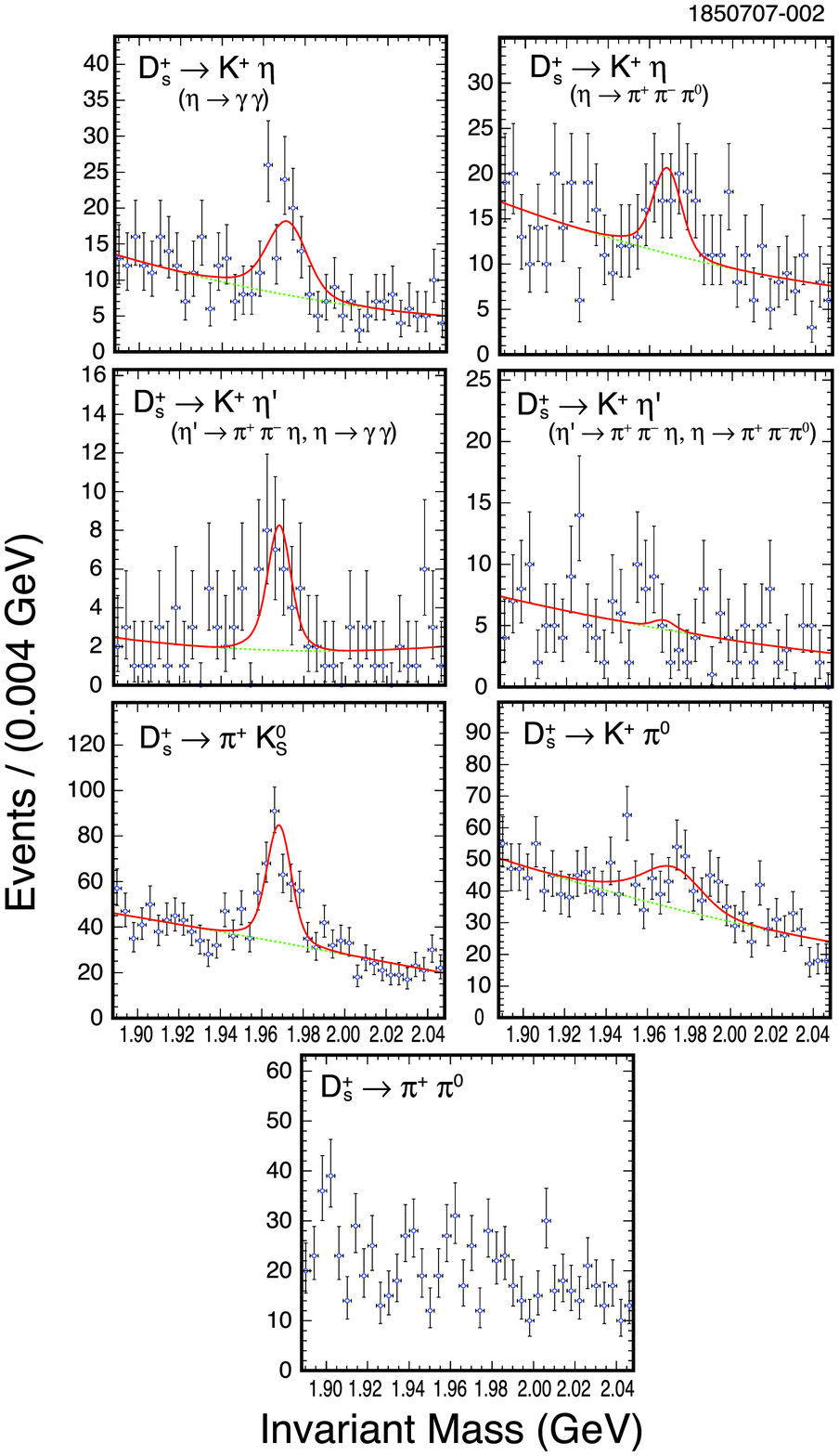}\\
  \caption{$M(D_s)$ distributions for Cabibbo-suppressed $D_s$ modes from
    data. The points are the data and the superimposed line is the
    fit (the dotted line is the fitted background)
    as described in the text. Also shown is the distribution for
    the isospin-forbidden decay $D^+_s \rightarrow \pi^+ \pi^0$.} 
  \label{fig:02}
\end{figure}

The resulting $M(D_s)$ distributions for the Cabibbo-favored and
Cabibbo-suppressed $D_s$ modes are shown in Figs.~\ref{fig:01} and
\ref{fig:02}, respectively. The points show the data and the lines are
fits. We perform a binned maximum likelihood fit (2~MeV bins) to
extract signal 
yields from the $M(D_s)$ distributions. For the signal, we use the sum
of two Gaussians for the line shape. The signal shape parameters are
determined by fits to $M(D_s)$ distributions obtained from
a GEANT-based Monte Carlo (MC) simulation~\cite{geant}. For the
background, we use a second-degree polynomial function, allowing the
overall scale, and the coefficient of the linear term relative to the
constant 
term, to float in the fits to the data. We constrain the (very small)
coefficient of the quadratic term relative to the constant term to the
value given by MC simulation. We include as a systematic error the
change in yield caused by varying the quadratic coefficient over a
reasonable range, typically doubling the quadratic term coefficient,
or setting it to zero. 
(For the favored modes, where the background is
relatively smaller, we allow the coefficient of the quadratic term to
float.)

\begin{table*}[tb]
  \begin{center}
    \caption{\label{tab:01}Observed yields from data and
      reconstruction efficiencies and their statistical
      uncertainties. 
      For the Cabibbo-suppressed modes, the statistical significance
      of the signal is also given (see the text for details). The
      efficiencies include sub-mode branching
      fractions~\cite{PDGValue}, and have been
      corrected to include several known small differences between
      data and Monte Carlo simulation.}
    \begin{tabular}{ l c l c rcl c rcl }
      \hline \hline
      $D_s$ Mode
      & ~~
      & Sub-Mode Decay
      & ~~
      & \multicolumn{3}{r}{Yield}~
      & ~~Significance ($\sigma$)~~
      & \multicolumn{3}{r}{Efficiency (\%)}  \\ \hline \hline

      $D^{+}_{s} \rightarrow \pi^+ \eta$ & & $\eta \rightarrow \gamma \gamma$ & & 908 & $\pm$ & 43 & & 9.97 & $\pm$ & 0.05 \\
      $D^{+}_{s} \rightarrow \pi^+ \eta$ & & $ \eta \rightarrow \pi^+ \pi^- \pi^0$ & & 512 & $\pm$ & 31 & & 5.00 & $\pm$ & 0.03 \\
      $D^{+}_{s} \rightarrow \pi^+ \eta'$ & & $\eta' \rightarrow \pi^+ \pi^- \eta, \eta \rightarrow \gamma \gamma$ & & 509 & $\pm$ & 25 & & 2.43 & $\pm$ & 0.02 \\
      $D^{+}_{s} \rightarrow \pi^+ \eta'$ & & $ \eta' \rightarrow \pi^+ \pi^- \eta, \eta \rightarrow \pi^+ \pi^- \pi^0$ & & 344 & $\pm$ & 24 & & 1.80 & $\pm$ & 0.01 \\
      $D^{+}_{s} \rightarrow K^+ K^{0}_{S}$ & & $K^{0}_{S} \rightarrow \pi^+ \pi^-$ & & 2174 & $\pm$ & 52 & & 26.13 & $\pm$ & 0.14 \\
      $D^{+}_{s} \rightarrow \pi^+ K^{0}_{S}$ & & $K^{0}_{S} \rightarrow \pi^+ \pi^-$ & & 206 & $\pm$ & 22 &12.9 & 29.93 & $\pm$ & 0.15 \\
      $D^{+}_{s} \rightarrow K^+ \pi^0$ & & $\pi^0 \rightarrow \gamma \gamma$ & & 129 & $\pm$ & 32 &4.7 & 30.90 & $\pm$ & 0.14 \\
      $D^{+}_{s} \rightarrow K^+ \eta$ & & $ \eta \rightarrow \gamma \gamma$ & & 68 & $\pm$ & 13 &5.6 & 8.93 & $\pm$ & 0.05 \\
      $D^{+}_{s} \rightarrow K^+ \eta$ & & $ \eta \rightarrow \pi^+ \pi^- \pi^0$ & & 45 & $\pm$ & 13 &3.7 & 4.39 & $\pm$ & 0.03 \\
      $D^{+}_{s} \rightarrow K^+ \eta'$ & & $ \eta' \rightarrow \pi^+ \pi^- \eta, \eta \rightarrow \gamma \gamma$ & & 25 & $\pm$ & 7 &5.7 & 2.10 & $\pm$ & 0.02 \\
      $D^{+}_{s} \rightarrow K^+ \eta'$ & & $ \eta' \rightarrow \pi^+ \pi^- \eta, \eta \rightarrow \pi^+ \pi^- \pi^0$ & & 3 & $\pm$ & 6 &0.0 & 1.53 & $\pm$ & 0.01 \\

      \hline \hline
    \end{tabular}
    
  \end{center}
\end{table*}

Results of the fits are shown in Table~\ref{tab:01}. Also given in
Table~\ref{tab:01} is the detection efficiency for each mode, and, 
for the Cabibbo-suppressed modes, the statistical significance of 
the signal. For $D^{+}_{s} \rightarrow K^+ \pi^0$, the statistical
significance is 4.7 standard deviations ($\sigma$), while for the
other modes
using $\eta \rightarrow \gamma \gamma$, the statistical significance
exceeds 5$\sigma$. The $\eta \rightarrow \pi^+ \pi^- \pi^0$ mode for 
$D^{+}_{s} \rightarrow K^+ \eta$ confirms the signal, at 3.7$\sigma$,
while for $D^{+}_{s} \rightarrow K^+ \eta'$, due to the very large
background of this mode, it gives no supporting evidence. 
For the $D^{+}_{s} \rightarrow \pi^+ K^{0}_{S}$ mode, the statistical
significance exceeds 10$\sigma$.
For all Cabibbo-favored modes, very clear signals are found in the
data.

We find no significant evidence for the isospin-forbidden decay 
$D^{+}_{s} \rightarrow \pi^+ \pi^0$, and therefore set an upper limit
on its rate. There is a large background from continuum events, and
Monte Carlo studies indicate that tightening the requirement on recoil
mass to $\pm 10$~MeV should improve the upper limit. The invariant
mass distribution with this requirement applied is shown in
Fig.~\ref{fig:02}. We apply a sideband subtraction to the invariant
mass distribution and obtain a yield of $17 \pm 25$
events. We interpret this result as implying a probability
distribution for the true number of events $N$ as a
Gaussian, centered on 17, with width $\sigma=25$, but truncated at
zero, so the probability distribution vanishes for a negative true
number of events. Ninety percent of the area of this distribution lies
below 52 events, which we take as the 90\% confidence level upper
limit on the true 
number of events. We normalize this upper limit on yield to that for 
$D^{+}_{s} \rightarrow K^+ K^{0}_{S}$, obtaining
$\mathcal{B}(D^{+}_{s} \rightarrow \pi^+ \pi^0)$ / 
$\mathcal{B}(D^{+}_{s} \rightarrow K^+ K^{0}_{S}) < 3.80 \times 10^{-2}$
(statistical only). Systematic errors, from the ratio of detection
efficiencies, are $\pm 1.8\%$ for the $K^0_S$, $\pm 4.2\%$ for the 
$\pi^0$, and other smaller errors, leading to a combined relative
systematic error of $\pm 5.1\%$. 
We conservatively increase the upper limit by 1.28
times the combined systematic errors, giving an upper
limit, including systematic errors, of $4.1 \times 10^{-2}$.

In principle, non-resonant $D_s$ decay could enter into our signal
modes with the same final particles. For example, non-resonant 
$D^+_s \rightarrow \pi^+ (\pi^+ \pi^- \pi^0)$ could appear in the 
$D^+_s \rightarrow \pi^+ \eta, \eta \rightarrow \pi^+ \pi^- \pi^0$ mode.
To understand the background
from non-resonant $D_s$ decay, we look at $M(D_s)$
distributions in the sideband region of the intermediate resonance 
($\eta$, $\eta'$, or $K^{0}_{S}$) invariant mass. 
These studies show that the non-resonant modes produce negligible
contributions to our signal modes.

\begin{table}
  \begin{center}
    \caption{\label{tab:02}Ratios of branching fractions
      of Cabibbo-suppressed modes
      to corresponding Cabibbo-favored modes. Uncertainties are
      statistical and systematic, respectively.}
    \begin{tabular}{l c c c c}
      \hline \hline
      Mode  
      & ~~
      & $\mathcal{B}_{\text{S}}/\mathcal{B}_{\text{F}}$$(10^{-2})$
      & 
      & 
      \\ \hline \hline
      $\mathcal{B}(D^{+}_{s} \rightarrow K^+ \eta)$ / $\mathcal{B}(D^{+}_{s} \rightarrow \pi^+ \eta)$   
      && 8.9 $\pm$ 1.5 $\pm$ 0.4 && \\
      $\mathcal{B}(D^{+}_{s} \rightarrow K^+ \eta')$ / $\mathcal{B}(D^{+}_{s} \rightarrow \pi^+ \eta')$   
      && 4.2 $\pm$ 1.3 $\pm$ 0.3 && \\
      $\mathcal{B}(D^{+}_{s} \rightarrow \pi^+ K^{0}_{S})$ / $\mathcal{B}(D^{+}_{s} \rightarrow K^+ K^{0}_{S})$
      && 8.2 $\pm$ 0.9 $\pm$ 0.2 && \\
      $\mathcal{B}(D^{+}_{s} \rightarrow K^+ \pi^0)$ / $\mathcal{B}(D^{+}_{s} \rightarrow K^+ K^{0}_{S})$
      && 5.0 $\pm$ 1.2 $\pm$ 0.6 && \\
      $\mathcal{B}(D^{+}_{s} \rightarrow \pi^+ \pi^0)$ / $\mathcal{B}(D^{+}_{s} \rightarrow K^+ K^{0}_{S})$
      && $< 4.1 $ (90\% CL) &&\\
      \hline \hline
    \end{tabular}
  \end{center}
\end{table}

For the modes with $\eta$ or $\eta'$ 
($\eta' \rightarrow \pi^+ \pi^- \eta$) in the final state, we
reconstruct these modes with $\eta$ decaying to $\gamma \gamma$ and to 
$\pi^+ \pi^- \pi^0$. For Cabibbo-favored modes, we
combine the two fit yields from the different $\eta$ decay modes
according to the fit yield fractional error. The weighting factors 
for both $D^{+}_{s} \rightarrow \pi^+ \eta$ 
and $D^{+}_{s} \rightarrow \pi^+ \eta'$
are 0.65 for $\eta \rightarrow \gamma \gamma$ 
and 0.35 for $\eta \rightarrow \pi^+ \pi^- \pi^0$. 
We apply the same weighting
factors to the corresponding Cabibbo-suppressed modes 
($D^{+}_{s} \rightarrow K^+ \eta$ and 
$D^{+}_{s} \rightarrow K^+ \eta'$). 
Doing so guarantees cancellation of systematic errors
between Cabibbo-favored and Cabibbo-suppressed modes. 
It also avoids a possible bias
that could come from using the errors on the Cabibbo-suppressed modes
to determine the weighting factors for them.

Ratios of branching fractions are computed for each of the
Cabibbo-suppressed modes and are presented in Table~\ref{tab:02}. They
are normalized with respect to the corresponding Cabibbo-favored modes. 
We use the $D^{+}_{s} \rightarrow K^+ K^0_S$ mode to normalize 
the $D^{+}_{s} \rightarrow K^+ \pi^0$ mode. The upper limit for the 
unobserved mode $D^{+}_{s} \rightarrow \pi^+ \pi^0$, normalized
with respect to $D^{+}_{s} \rightarrow K^+ K^0_S$, is also shown in
Table~\ref{tab:02}.

We have considered several sources of systematic uncertainty.
Finite MC statistics in determining reconstruction efficiencies
introduces uncertainties at the level of less than 1\%.
The uncertainty associated with the efficiency for
finding a track is 0.3\%; an additional 0.6\% systematic
uncertainty for each kaon track is added.
The relative systematic uncertainties for
$\pi^0$ and $K^0_S$ efficiencies are 4.2\% and  
1.8\%, respectively. The systematic uncertainty for $\eta$
efficiencies cancels in all ratios.
Uncertainties in the charged pion and kaon identification
efficiencies are 0.3\% per pion and 1.3\% per
kaon~\cite{HadronPaper}. 
The systematic uncertainties from the $K^0_S$ flight significance
requirement and the low-momentum track veto are 0.5\% and 0.3\%, respectively.
The signal shape parameters are taken from MC simulation, and have
uncertainties related to possible flaws in simulation. We
estimate this systematic uncertainty by allowing the signal shape
parameters of the favored modes to float, and having the parameters of
the suppressed modes track those of the favored modes. We find
uncertainties of 0.2\% to 2.9\%, depending on mode. 
For the suppressed modes, the background
quadratic term is also taken from MC simulation. 
We vary that term over a reasonable range, finding a systematic
error of 2.4\% to 10.6\%, depending on mode.

In calculating the relative systematic uncertainties for the measured
ratio of Cabibbo-suppressed mode branching fractions to
Cabibbo-favored mode branching fractions 
($\mathcal{B}_{\text{Suppressed}}/\mathcal{B}_{\text{Favored}}$),
cancellation of uncertainties has been taken into account. 
The systematic uncertainties that do not
cancel in the ratios are added in quadrature
to obtain the total systematic uncertainties shown as the second error
in Table~\ref{tab:02}. For the upper limit in
Table~\ref{tab:02}, the systematic uncertainties have been included as
previously described. Systematic uncertainties
for all measured ratios are at most half the statistical uncertainties.

\begin{table}
  \begin{center}
    \caption{\label{tab:03}Measured $CP$ asymmetries in
      Cabibbo-suppressed decay modes. Only statistical
      uncertainties are included. Systematic errors are negligible by
      comparison.}
    \begin{tabular}{l c r c l}
      \hline \hline
      Mode  
      & ~~~~~~~~~~~~~~~~
      &\multicolumn{3}{r}{$(\mathcal{B}_{+} - \mathcal{B}_{-}) / (\mathcal{B}_{+} + \mathcal{B}_{-})$(\%)}
      \\ \hline \hline
      $\mathcal{A}(D^{+}_{s} \rightarrow K^+ \eta)$  
      &&~~~~~~~~ $-20$ &$\pm$& 18 \\
      $\mathcal{A}(D^{+}_{s} \rightarrow K^+ \eta')$
      && $-17$ &$\pm$& 37 \\
      $\mathcal{A}(D^{+}_{s} \rightarrow \pi^+ K^{0}_{S})$
      && 27  &$\pm$& 11 \\
      $\mathcal{A}(D^{+}_{s} \rightarrow K^+ \pi^0)$
      && 2   &$\pm$& 29\\
      \hline \hline
    \end{tabular}
  \end{center}
\end{table}

The Standard Model predicts that direct $CP$ violation in $D$ decays,
{\it e.g.}, a difference in the branching fractions for 
$D^{+}_{s} \rightarrow K^+ \eta$ and $D^{-}_{s} \rightarrow K^- \eta$,
will be vanishingly small. As a search for evidence 
of non-Standard-Model physics, we have therefore measured the $CP$
asymmetries  
$\mathcal{A} \equiv (\mathcal{B}_{+} - \mathcal{B}_{-}) / (\mathcal{B}_{+} + \mathcal{B}_{-})$
for the four Cabibbo-suppressed $D_s$ decay modes we are studying. 
Results are given in Table~\ref{tab:03}. Errors shown are
statistical. The systematic errors, from the differences in efficiency for
detecting $K^+$~{\it vs.}~$K^-$ and $\pi^+$~{\it vs.}~$\pi^-$, are
$< 2.0\%$, negligible by comparison. All asymmetries are consistent
with zero.

In summary, we report first observations of four Cabibbo-suppressed
decays of $D_s$ mesons, and measure the ratio of their
branching fractions to the corresponding Cabibbo-favored modes. We
find those ratios to be of order $|V_{cd}/V_{cs}|^{2} \approx 1 /20$
in agreement with naive expectations. We report a first upper limit on
the isospin-forbidden decay $D^{+}_{s} \rightarrow \pi^+ \pi^0$. The $CP$
asymmetries for the four Cabibbo-suppressed decays are consistent with
zero, as predicted by the Standard Model.

We gratefully acknowledge the effort of the CESR staff
in providing us with excellent luminosity and running conditions.
This work was supported by
the A.P.~Sloan Foundation,
the National Science Foundation,
the U.S. Department of Energy, and
the Natural Sciences and Engineering Research Council of Canada.


\begin{thebibliography}{99}

\bibitem{RosnerPaper}   
  C.~W.~Chiang, Z.~Luo, and J.~L.~Rosner, Phys. Rev. {\bf D 67}, 014001 (2003).
\bibitem{PDGValue}      
  W.-M. Yao $et ~ al$. (Particle Data Group),
  J. Phys. G \textbf{33}, 1 (2006).
\bibitem{cleo_detector} 
  Y.~Kubota {\it et~al.}, {Nucl. Instrum. Meth. Phys. Res., Sect. A} \textbf{320}, {66} ({1992});
  D.~Peterson {\it et~al.}, {Nucl. Instrum. Meth. Phys. Res., Sect. A} \textbf{478}, {142} ({2002});
  M.~Artuso {\it et~al.}, {Nucl. Instrum. Meth. Phys. Res., Sect. A} \textbf{554}, {147} ({2005}).
\bibitem{DataSample}
  R.~Poling,
  {\it In the Proceedings of 4th Flavor Physics and CP Violation 
    Conference (FPCP 2006), Vancouver, British Columbia, Canada, 9-12 Apr 
    2006, pp 005}
  [arXiv:hep-ex/0606016].
\bibitem{HadronPaper}  
  CLEO Collaboration, Q.~He~{\it et~al.}, Phys. Rev. Lett. {\bf 95}, 121801~(2005).
\bibitem{geant}        
  R.~Brun {\it et~al.}, {\tt GEANT 3.21}, CERN Program Library Long Writeup W5013, unpublished.

\end{thebibliography}
\end{document}